\documentclass[10pt,showpacs,floatfix,letterpaper,amsmath,amsfonts,amssymb,aps,prl,reprint,superscriptaddress]{revtex4-1}
\usepackage[latin1]{inputenc}
\usepackage{bm}
\usepackage{graphicx}
\usepackage{tensor}
\usepackage{enumerate}

\newcommand{\ket}[1]{\lvert #1\rangle}
\newcommand{\bra}[1]{\langle #1\rvert}
\newcommand{\braket}[2]{\langle #1 \vert #2 \rangle}
\newcommand{\pr}[1]{\ket{#1}\bra{#1}}

\newcommand{\abs}[1]{\left| #1 \right|}
\newcommand{\op}[1]{\hat{#1}}

\begin{document}
\title{Violating the Modified Helstrom Bound with Nonprojective Measurements}
\author{Justin Dressel}
\affiliation{Department of Electrical and Computer Engineering; University of California, Riverside, CA 92521, USA}
\author{Todd A. Brun}
\affiliation{Communication Sciences Institute, University of Southern California, Los Angeles, CA 90089, USA}
\author{Alexander N. Korotkov}
\affiliation{Department of Electrical and Computer Engineering; University of California, Riverside, CA 92521, USA}

\date{\today}

\begin{abstract}
  We consider the discrimination of two pure quantum states with three allowed outcomes: a correct guess, an incorrect guess, and a non-guess.  To find an optimum measurement procedure, we define a tunable cost that penalizes the incorrect guess and non-guess outcomes.  Minimizing this cost over all projective measurements produces a rigorous cost bound that includes the usual Helstrom discrimination bound as a special case.  We then show that nonprojective measurements can outperform this \textit{modified Helstrom bound} for certain choices of cost function.  The Ivanovic-Dieks-Peres unambiguous state discrimination protocol is recovered as a special case of this improvement.  Notably, while the cost advantage of the latter protocol is destroyed with the introduction of any amount of experimental noise, other choices of cost function have optima for which nonprojective measurements robustly show an appreciable, and thus experimentally measurable, cost advantage. Such an experiment would be an unambiguous demonstration of a benefit from nonprojective measurements.
\end{abstract}


\maketitle

A fundamental consequence of quantum mechanics is the inability to perfectly distinguish between two non-orthogonal quantum states.  Any attempt to guess which state is which after making a measurement will have an unavoidable probability of error that is bounded from below, as shown originally by Helstrom \cite{Helstrom1967,Helstrom1968} and in related work by Holevo \cite{Holevo1973}. This lower bound, known as the Helstrom bound (HB), grows with the overlap of the two states being discriminated.

The HB can be circumvented, however, if a third option is added to the guessing game.  Ivanovic, Dieks, and Peres showed that if one can also decline to guess after a measurement, then it is possible to reduce the probability of error to zero while still retaining a significant chance of guessing correctly \cite{Ivanovic1987,Dieks1988,Peres1988}. Intriguingly, to maximize the correct guess probability in such ``Unambiguous State Discrimination'' (USD), it is not sufficient to use standard projective measurements; instead, one must use generalized (\emph{nonprojective}) measurements \cite{Davies1976,Kraus1983}.

This advantage of nonprojective measurements in state discrimination is so surprising that it has become a featured example in modern quantum information textbooks (e.g., \cite{Nielsen2000,Barnett2010}), and has led to considerable research, both in theory \cite{Yuen1975,Chefles1998,Chefles1998b,Sun2002,Eldar2003,Rudolph2003,Raynal2003,Feng2004,Eldar2005,Croke2006,Bergou2007,Bae2013,Chefles1998c,Zhang1999,Fiurasek2003,Touzel2007,Andersson2012} and experiment \cite{Huttner1996,Barnett1997b,Clarke2001,Mohseni2004,Barnett2004,Mosley2006,Wittmann2008,Lu2010,Steudle2011,Becerra2013} (reviewed, e.g., in \cite{Barnett2009,Bergou2010}).  Most of this work has focused on the extreme cases of zero declining (as with the HB) or zero error (as with USD), with fewer papers considering intermediate cases that minimize the declining probability given a fixed nonzero error rate \cite{Chefles1998c,Zhang1999,Fiurasek2003,Touzel2007,Andersson2012}. Moreover, to our knowledge all but one \cite{Andersson2012} of these few works have neglected the effect that experimental imperfections will have upon the accessible minima. We are thus not aware of any paper that discusses a rigorous bound suitable to experimentally demonstrate that nonprojective measurements have a definitive advantage over projective measurements.

In this paper, we provide such a bound by considering a simple cost function that interpolates between the HB and USD extremes as special cases. This approach allows us to formulate a rigorous lower bound for this cost that is accessible to projective measurements, and then demonstrate that nonprojective measurements can violate this bound.  Realistic experimental noise changes the theoretical minimum for each choice of cost, affecting which violations can be observed.  Notably, the cost advantage of USD is completely destroyed with any amount of experimental noise.  Nevertheless, we show that nonprojective measurements still show an appreciable advantage for intermediate cost functions, making this advantage experimentally accessible to current implementations of generalized measurements, including experiments with superconducting qubits \cite{Katz2006,Weber2014,Hatridge2013,Groen2013,Barends2014,Dressel2014}.

\emph{State Discrimination}.---
Consider the following game:  An agent, whom we shall name Alice, prepares one of two pure quantum states with equal probability,
\begin{align}
  \ket{\psi_0} &= \ket{0}, &
  \ket{\psi_1} &= \cos\theta\,\ket{0} + \sin\theta\,\ket{1},
\end{align}
and sends it to another agent, Bob, who wishes to determine which state Alice has prepared.  To write these states, we have used the fact that any two states lie in a plane that can be spanned by two orthogonal vectors, which we label $\ket{0}$ and $\ket{1}$.  These states form a basis for an effective qubit, even though the implementation Hilbert space may have more dimensions.

Once he has obtained a state from Alice, Bob is allowed to measure it in any way that he pleases, after which he must either guess the state or decline to guess.   There are thus three possible results for a single trial of this game: (1) Bob can guess the state correctly, (2) Bob can guess the state incorrectly, or (3) Bob can decline to guess.  Hence, if Bob uses a consistent measurement strategy for many trials, three probabilities will emerge that correspond to these results: (1) correctly guessing with probability $p_c$, (2) wrongly guessing with probability $p_w$, and (3) declining to guess with probability $p_d$.  These probabilities will satisfy $p_c + p_w + p_d = 1$.

To quantify how well Bob is playing the game, Alice defines a suitable cost function that penalizes the unfavorable outcomes (i.e., $p_w$ and $p_d$) in some proportion.  The simplest linear cost function has the general form
\begin{equation}\label{eq:cost}
  C = w\, p_w + d\, p_d,
\end{equation}
where $w$ and $d$ are positive weights that penalize incorrect guesses and non-guesses, respectively.  

For simplicity of discussion, in most of what follows we will normalize the cost function by the weight $w$ (i.e., $C\to C/w$), to leave only a single parameter $k$,
\begin{align}\label{eq:weights}
  C &= p_w + k\,p_d, & k &= d/w,
\end{align}
that indicates the penalty for not guessing relative to that of incorrectly guessing.  To analyze the limit $w\to \infty$, we can use the modified cost $C/k=p_w/k +p_d$.

In terms of the single parameter $k$, we have the following limiting behaviors: (i) When $k\to\infty$, non-guesses are intolerable, so the minimized cost effectively reduces to $\min(C) =\min(p_{w})$, subject to the constraint $p_d = 0$.  This limiting case corresponds to the standard two-outcome discrimination game \cite{Helstrom1967,Helstrom1968,Holevo1973}, so the minimized cost will be equal to the usual HB.  (ii) When $k\to 0$, there is no penalty for non-guesses, so it is always better to decline ($p_d = 1$) to produce $\min (C) \to 0$.  However, after rescaling to $C/k$ the limit $k\to 0$ is non-trivial: Then wrong guesses become intolerable, so the minimized cost reduces to $\min (C/k) =\min (p_d)$ subject to the constraint $p_w = 0$, which corresponds to the USD game \cite{Ivanovic1987,Dieks1988,Peres1988}.

We see that our formulation of the state discrimination game with a linear cost function is sufficiently general to contain both the HB and USD games as special cases at extremes of $k$.  We are thus particularly interested in the optimal strategies for cost functions between these well-known extremes.  For intermediate $k$, we analyze when nonprojective quantum measurements are advantageous compared with projective measurements, and find the size of this advantage under realistic experimental conditions.

\emph{Modified Helstrom Bound}.---
We first find a rigorous lower bound for the cost function \eqref{eq:cost} if only projective measurements are allowed within the qubit space. Nonprojective measurements will be able to violate this bound.  Note that a projective measurement of a qubit fully determines the post-measurement state, so an additional measurement would not bring additional information. 
Therefore, there are only two possible optimal strategies for discriminating two pure states:
\begin{enumerate}[(a)]
  \item Always guess both states.  That is, perform one projective measurement in an orthogonal basis $\{ \ket{\phi_0}, \ket{\phi_1} \}$, identifying $\ket{\phi_0}$ as a guess of $\ket{\psi_0}$ and the orthogonal state $\ket{\phi_1}$ as a guess of $\ket{\psi_1}$.
  \item Only guess one state.  That is, perform one projective measurement in an orthogonal basis $\{\ket{\phi_0}, \ket{\phi_1} \}$, with $\ket{\phi_0}$ used as a guess of $\ket{\psi_0}$, while treating $\ket{\phi_1}$ as a non-guess outcome.
\end{enumerate}
Other intermediate strategies that probabilistically combine these two will not be optimal due to the convexity of the linear cost function. Trivial state exchanges $0\leftrightarrow 1$ give the same performance.

For strategy (a) the game probabilities are $p_d = 0$ and
\begin{align}
  p_w &= \frac{\abs{\braket{\phi_0}{\psi_1}}^2 + \abs{\braket{\phi_1}{\psi_0}}^2}{2} = \bra{\phi_0}\op{A}\ket{\phi_0},
\end{align}
where $\op{A} = \left(\hat{1} + \pr{\psi_1} - \pr{\psi_0}\right)/2$ and the factor of $1/2$ indicates the 50:50 preparation probability for each state $\ket{\psi_i}$.  The minimum $p_w$ is the minimum eigenvalue of $\op{A}$, so the minimum cost in Eq.~\eqref{eq:cost} for strategy (a) is this eigenvalue scaled by $w$:
\begin{equation}\label{eq:hb}
  C_{\min}^{(a)} = w\,\left(1 - \abs{\sin\theta}\right)/2.
\end{equation}
The weight $w$ vanishes when using the normalization of Eq.~\eqref{eq:weights}, so the cost reduces to the usual HB \cite{Helstrom1967,Helstrom1968}.

For strategy (b) the game probabilities are
\begin{align}
  p_w &= \frac{\abs{\braket{\phi_0}{\psi_1}}^2}{2}, & p_d &= \frac{\abs{\braket{\phi_1}{\psi_0}}^2 + \abs{\braket{\phi_1}{\psi_1}}^2}{2},
\end{align}
so the minimum cost is the minimum eigenvalue of the operator $\op{B} = w\,\pr{\psi_1}/2 + d\,\left[\op{1} - \left(\pr{\psi_0} + \pr{\psi_1}\right)/2 \right]$, which is
\begin{align}
  C_{\min}^{(b)} &= \frac{w + 2d}{4} - \sqrt{\left[\frac{w - 2d}{4}\right]^2 + \frac{d\left[w-d\right]}{4}\,\sin^2\theta}.
\end{align}
For the normalization of Eq.~\eqref{eq:weights}, this cost simplifies to $C_{\min}^{(b)} = [1+2k - \sqrt{1-2k(1-k)(1+\cos2\theta)}]/4$.

\begin{figure}[tbh!]
  \includegraphics[width=\columnwidth]{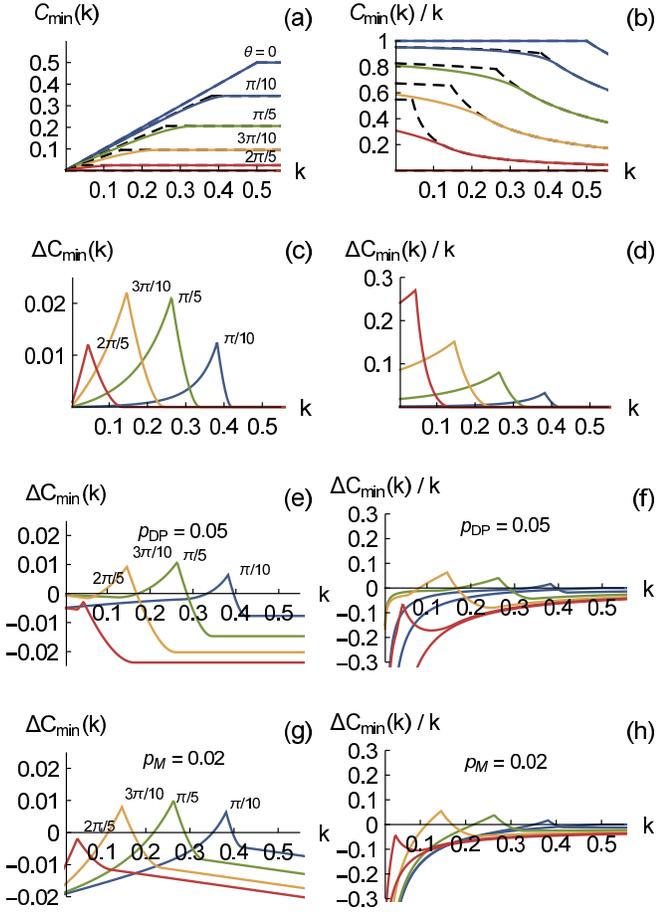}
  \caption{(a) Minimum cost for discriminating $\ket{\psi_0} = \ket{0}$ and $\ket{\psi_1} = \cos\theta\,\ket{0}+\sin\theta\,\ket{1}$ by a measurement, using the cost function $C(k) = p_w + k\,p_d$ to penalize wrong, $p_w$, and declined, $p_d$, guess probabilities.  Shown are separation angles $\theta$ in increments of $\pi/10$. Dashed lines show the modified Helstrom bound $C_{\rm MH}$, Eq.\ (\ref{eq:MH}), attainable with a projective measurement, with the horizontal part being the usual Helstrom bound.  Solid lines (same colors in all panels) show the minimum cost $C_{\rm min}$ for nonprojective measurements, and violate the MH bound for a range of $k$.
  (b) Same as in (a) for the scaled cost function $C(k)/k$.  This scaling permits the case of unambiguous state discrimination to be recovered at $k\to 0$.  (c) Violation of the MH bound, $\Delta C_{\rm min}=C_{\rm MH}-C_{\rm min}$, showing the difference between the dashed and solid curves in (a).  (d) Violation of the scaled bound,  $\Delta C_{\rm min}/k$. (e) and (f) Same as (c,d) but with 5\% probability $p_{DP}$ of depolarizing the states.  (g) and (h) Same as (c,d) but with 2\% probability $p_M$ of misidentifying the measured result.  These imperfections increase $C_{\rm min}$ and therefore decrease the violation $\Delta C_{\rm min}$. Note that for the USD case [$k\to 0$ in (f) and (h)] the cost advantage is fully destroyed by experimental imperfections, but the MH bound violation is still possible for intermediate $k$.}
  \label{fig:modhelsimple}
\end{figure}

The minimum cost of these two possible strategies is the best that Bob can do using projective measurements; we will call it the \emph{modified Helstrom} (MH) bound
\begin{equation}
  C_{\text{MH}} = \min\{C_{\min}^{(a)},\,C_{\min}^{(b)}\}.
\label{eq:MH}\end{equation}
This bound [with the normalization of Eq.~\eqref{eq:weights}] as a function of $k$ is illustrated with dashed lines in Figs.~\ref{fig:modhelsimple}(a) and \ref{fig:modhelsimple}(b) for various choices of the separation angle $\theta$ between $\ket{\psi_0}$ and $\ket{\psi_1}$.  Each kink indicates a switch between the two projective strategies where $C_{\min}^{(a)} = C_{\min}^{(b)}$.  As we discuss later, nonprojective measurements maximally violate the MH bound at precisely these critical (optimal) values of $k$, which depend on the separation angle $\theta$,
\begin{align}\label{eq:kmax}
  k_{\text{opt}}(\theta) &= \frac{1}{2}\left[1 + \frac{\sqrt{1+3\cos^2\theta} - 2}{\abs{\sin\theta}}\right].
\end{align}

\emph{Nonprojective Measurements}.---
Unlike projective strategies that can have only two physical outcomes, nonprojective measurements can naturally use three physical outcomes for the three choices in the discrimination game. This is what permits nonprojective measurements to have an advantage over projective measurements. (It is simple to show that using four or more physical outcomes will not lead to further improvement.)

Without loss of generality, we consider a concrete implementation of a three-outcome nonprojective strategy as a cascade of two binary-outcome measurements, the first being a partial projection (see, e.g., \cite{Dressel2014}) and the second being a full projection. (For optimal cascaded strategies, the second measurement will always be projective so that it extracts the maximum remaining information.) The advantage of using this cascading strategy is a relatively easy implementation with existing experimental qubit architectures, especially with superconducting qubits \cite{Katz2006,Weber2014,Hatridge2013,Groen2013,Barends2014}.

To implement the three-outcome cascade, Bob uses the following procedure:
\begin{enumerate}[(a)]
  \item Measure in a basis that includes the state $\ket{\phi^{(1)}_0} = \cos\varphi_1 \ket{0} + \sin\varphi_1 \ket{1}$,  with a strength $s\in[0,1]$ (see, e.g., \cite{Dressel2014}).  If the outcome $\ket{\phi^{(1)}_0}$ is obtained, treat this as a guess of $\ket{\psi_0}$.
  \item Otherwise, perform a second projective measurement in a basis that includes the state $\ket{\phi^{(2)}_1} = \cos\varphi_2 \ket{0} + \sin\varphi_2 \ket{1}$.  If the outcome $\ket{\phi^{(2)}_1}$ is obtained, treat this as a guess of $\ket{\psi_1}$.
  \item The remaining outcome is treated as a non-guess.
\end{enumerate}
Note that we omit relative phases in both measurement bases above, since optimal measurements of any strength will always be in the same plane as the two states being discriminated.

The three possible measurement outcomes of this cascade then correspond to the following partial projection operators \cite{Nielsen2000,Dressel2014} that are parametrized by the two angles $\varphi_1,\varphi_2\in[-\pi,\pi]$, as well as the strength $s\in[0,1]$:
\begin{subequations}
\begin{align}
  \op{M}_0 &= s \pr{\phi^{(1)}_0}, \\
  \op{M}_1 &= \pr{\phi^{(2)}_1}\,\sqrt{\op{1} - s^2\pr{\phi^{(1)}_0}}, \\
  \op{M}_d &= \sqrt{\op{1}-\pr{\phi^{(2)}_1}}\, \sqrt{\op{1} - s^2\pr{\phi^{(1)}_0}}.
\end{align}
\end{subequations}
These operators satisfy the usual completeness  condition $\op{M}_0^\dagger \op{M}_0 + \op{M}_1^\dagger \op{M}_1 + \op{M}_d^\dagger \op{M}_d = \op{1}$, and produce the game probabilities:
\begin{subequations}
\begin{align}
  p_c &=  \frac{1}{2}\left( \bra{\psi_0}\op{M}_0^\dagger\op{M}_0\ket{\psi_0} + \bra{\psi_1}\op{M}_1^\dagger\op{M}_1\ket{\psi_1}\right), \\
  p_w &= \frac{1}{2}\left( \bra{\psi_0}\op{M}_1^\dagger\op{M}_1\ket{\psi_0} + \bra{\psi_1}\op{M}_0^\dagger\op{M}_0\ket{\psi_1}\right), \\
  p_d &= \frac{1}{2}\left( \bra{\psi_0}\op{M}_d^\dagger\op{M}_d\ket{\psi_0} + \bra{\psi_1}\op{M}_d^\dagger\op{M}_d\ket{\psi_1}\right).
\end{align}
\end{subequations}
With the strength $s=1$ this cascading implementation can reproduce either of the projective strategies considered before, thus recovering the MH bound $C_{\text{MH}}$ when projections are indeed optimal.

To find the minimum cost, as well as the optimum parameters $(\varphi_1,\varphi_2,s)$, we numerically minimize the cost function in Eq.~\eqref{eq:weights} for each $k$ independently, using a Nelder Mead optimization algorithm.  In Fig.~\ref{fig:modhelsimple}(a) we show the resulting minimum cost $C_{\rm min}$ for each $k$ as the solid curves. For each separation angle $\theta$, there is a certain value $k_{\rm HB}(\theta)\leq 1/2$, above which the usual Helstrom bound in Eq.~\eqref{eq:hb} is recovered (the horizontal part of the line): in this regime projective measurements are the optimal strategy. However, for $0<k<k_{\rm HB} (\theta)$ the nonprojective measurements \emph{violate} the HB as well as the MH bound (dashed lines).  Using the results in Ref.~\cite{Chefles1998c}, we also derive the analytic form of the ideal minimum cost in this range (which coincides with the numerical results)
\begin{align}
  C_{\rm min} &= k[k - (1-k)\cos\theta]/(2k-1).
\end{align}
For $k \geq k_{\rm HB}$, $C_{\rm min}$ is the HB in Eq.~\eqref{eq:hb} (with $w=1$).

\begin{figure}[tb]
  \includegraphics[width=\columnwidth]{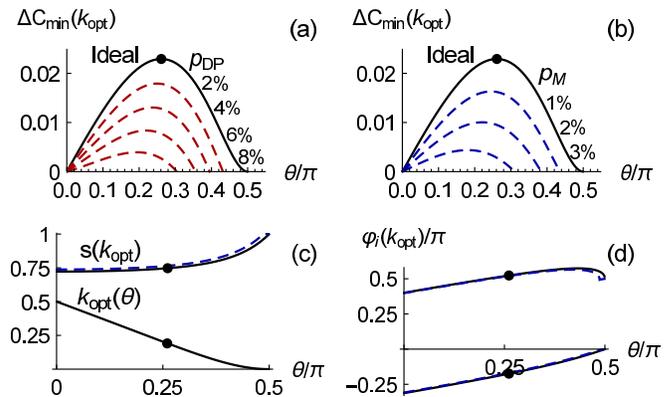}
  \caption{(a) Maximum violation of the modified Helstrom bound as a function of the state-separation angle $\theta$.  The ideal violation (solid line) is reduced in the presence of depolarization decoherence with strength $p_{DP}$ increasing in increments of $2\%$. (dashed lines).  (b) The ideal violation (solid line) is similarly reduced in the presence of measurement misidentification errors with probability $p_M$ increasing in increments of $1\%$.  (c) Lower curve: the optimal cost parameter $k_{\rm opt}(\theta)$ for the maximum MH bound violation [Eq.\ (\ref{eq:kmax}), peaks in Fig.\ 1(c)] in the ideal case.  Upper curves: the optimal measurement strength $s$ for the cascaded partial measurement scheme, as a function of $\theta$, for the optimal cost parameter $k_{\rm opt}(\theta)$.  (d) Optimal angles $\varphi_1$ (lower curves) and $\varphi_2$ (upper curves) for the cascaded partial measurement scheme.   In both (c) and (d) the solid curves are for the ideal case, while the (almost identical) dashed curves include 2\% misidentification noise.  The black dots indicate the globally maximum ideal violation.
  }
  \label{fig:modhel4simple}
\end{figure}

The maximum violation for each $\theta$ is shown in Figs.~\ref{fig:modhel4simple}(a) and (b), and occurs at the MH bound kinks $k_{\rm opt}(\theta)$ given by Eq.~\eqref{eq:kmax} [lower curve in Fig.~\ref{fig:modhel4simple}(c)]. The values of the optimal parameters $(\varphi_1,\varphi_2,s)$ minimizing the cost at these kinks are shown in Figs.~\ref{fig:modhel4simple}(c) and (d) as the solid lines.  We also show the result for minimizing the scaled cost $C/k$ in Fig.~\ref{fig:modhelsimple}(b), which recovers the special case of USD in the limit $k\to 0$.  For visual clarity, we show the cost improvement $\Delta C_{\rm min} = C_{MH}-C_{\rm min}$ (the difference between the MH bound and minimized cost) in Figs.~\ref{fig:modhelsimple}(c) and \ref{fig:modhelsimple}(d).

\emph{Experimental Imperfections}.---
We use two simple models to describe experimental imperfections, which may reduce or fully destroy the possible violations of the MH bound. First, we model decoherence with the depolarization process, assuming that the initial state prepared by Alice is replaced by the fully mixed state with a probability $p_{DP}$.  Second, we assume that the binary readouts of the cascaded measurement can be misidentified with an error probability $p_M$, which introduces spurious identification noise into the game.

In Figs.~\ref{fig:modhelsimple}(e) and (f) we show the effect of depolarization decoherence with $p_{DP} = 0.05$ for the optimized cost improvement.  Similarly, in (g) and (h) we show the effect of adding misidentification noise with $p_M = 0.02$.  These plots can be compared to the ideal results in Figs.~\ref{fig:modhelsimple}(c) and (d). Both types of imperfections have a similar effect on the maximum violations (with more sensitivity to $p_{M}$ than to $p_{DP}$), despite different dependences on $k$.

The cost improvement for USD case ($k\to 0$) in Fig.~\ref{fig:modhelsimple}(d) is completely destroyed for any $p_{DP}>0$ or $p_M > 0$, making this well-known protocol actually worse than ideal projective measurements for any experimentally realistic implementation of the state discrimination game.  Nevertheless, the globally maximum cost improvement, shown in Fig.~\ref{fig:modhel4simple}(a) and (b), only decreases approximately linearly as either $p_{DP}$ or $p_M$ increase.  Even with these imperfections, nonprojective measurements show an improvement over projective measurements around the critical parameter values $k_{\rm opt}(\theta)$.  The angles and strength associated with these maximum cost improvements including misidentification noise with $p_M=0.02$ are shown in Figs.~\ref{fig:modhel4simple}(c) and (d) as the dashed lines, which do not significantly differ from the ideal values. The MH bound violation requires $p_{DP}<0.101$ and $p_M<0.041$. 

\emph{Conclusion}.---
We have considered the two-state three-outcome discrimination game using a simple linear cost function to penalize the unfavorable outcomes. The original Helstrom discrimination problem, as well as the unambiguous state discrimination of Ivonovic, Dieks, and Peres are recovered as special cases.  Minimizing the cost function using only projective measurements produces what we name the modified Helstrom bound.

Nonprojective measurements can violate this modified bound. Notably, for cost functions intermediate between the well-studied extremes, the violations are robust against the introduction of (small) experimental imperfections.  In contrast, the cost advantage of the unambiguous state discrimination is completely destroyed with the addition of any amount of noise.  An experimental demonstration of modified Helstrom bound violations would require less than $\sim$10\% decoherence and $\sim$4\% readout error, making it a stringent-but-accessible test for modern quantum computing implementations.

\emph{Acknowledgments}.---
We thank Eyob Sete for valuable comments.  The research was funded by the Office of the Director of National Intelligence (ODNI), Intelligence Advanced Research Projects Activity (IARPA), through the Army Research Office (ARO) Grant No. W911NF-10-1-0334. All statements of fact, opinion, or conclusions contained herein are those of the authors and should not be  construed as representing the official views or policies of IARPA, the ODNI, or the U.S. Government. We also acknowledge support from the ARO MURI Grant No. W911NF-11-1-0268.

\end{document}